\documentstyle[amstex,amssymb,twocolumn,aps,epsfig]{revtex}

\twocolumn

\begin{document}
\draft
\title{Temporal build-up of electromagnetically induced transparency and absorption
resonances in degenerate two-level transitions}
\author{P. Valente, H. Failache and A. Lezama}
\address{Instituto de F\'{\i}sica, Facultad de Ingenier\'{\i}a. Casilla de correo 30. \\
11000, Montevideo, Uruguay.}
\date{\today }
\maketitle

\begin{abstract}
The temporal evolution of electromagnetically induced transparency (EIT) and
absorption (EIA) coherence resonances in pump-probe spectroscopy of
degenerate two-level atomic transition is studied for light intensities
below saturation. Analytical expression for the transient absorption spectra
are given for simple model systems and a model for the calculation of the
time dependent response of realistic atomic transitions, where the Zeeman
degeneracy is fully accounted for, is presented. EIT and EIA resonances have
a similar (opposite sign) time dependent lineshape, however, the EIA
evolution is slower and thus narrower lines are observed for long
interaction time. Qualitative agreement with the theoretical predictions is
obtained for the transient probe absorption on the $^{85}Rb$ $D_{2}$ line in
an atomic beam experiment.
\end{abstract}

\pacs{42.50.Gy, 42.50.Md, 32.80.Bx, 32.70.Jz.}

\preprint{}

\section{Introduction.}

It is well known that the interaction of an atomic system with correlated
optical waves can lead to quantum coherence in the atomic state. In turn,
the atomic coherence dramatically modifies the interaction with the light 
\cite{SCULLYBOOK}. A good example is provided by the effect of
electromagnetically induced transparency (EIT) \cite{HARRIS97} where an
absorbing medium can be rendered transparent to a probe field if it is
simultaneously submitted to the action of a coupling field. EIT takes place
when the two fields satisfy a two photon resonance condition between atomic
levels. The corresponding (coherence) resonance for the probe absorption is
called a ``dark resonance'' and is associated to the system being driven
into a ``dark state'' i.e. a coherent superposition of atomic states
uncoupled to the light field\cite{AGAPEV93,ARIMONDOREV}.

In recent years, a new kind of coherence resonance was observed in which the
atomic coherence results in an increase in the absorption of a probe field 
\cite{AKULSHIN98}. This phenomenon designated electromagnetically induced
absorption (EIA), bears except for the sign of the resonance, several common
features with EIT although it cannot be associated to the system being
driven into a particular superposition of atomic states. EIA was first
observed on a transition between two atomic levels with angular momentum
degeneracy \cite{AKULSHIN98}. The degenerate two-level system (DTLS) was
illuminated by \ a resonant pump field and probed with a tunable probe
field. It was clearly established that EIA is essentially a multilevel
effect in the sense that it only takes place if both the lower and the upper
atomic levels posses degeneracy. The condition for the occurrence of EIA in
atomic transitions is $F_{e}>F_{g}>0$ where $F_{e}$ and $F_{g}$ are the
total angular momenta of the excited and ground state respectively\cite
{AKULSHIN98,LEZAMA99,LEZAMA00,DANCHEVA00}. The simplest model system for
which EIA\ occurs is a four-level $N$ system theoretically investigated by
Taichenachev {\it et al} \cite{TAICHENACHEV00}. The analysis of this system
reveals that EIA is the consequence of the transfer of coherence from the
upper to the lower atomic levels via spontaneous emission.

The transient properties of EIT has been explored by several authors in
simple model systems \cite{AGAPEV93}. In \cite{LI95} the transient behavior
of EIT in three level system when the pump field is suddenly switched on is
considered and the onset of Rabi oscillations for saturating pump field
intensities discussed. The occurrence of EIT for low intensities was
analyzed in \cite{JYOTSNA95} for a closed three level $\Lambda $ system. It
is established that EIT can take place even for very low intensities and
that the characteristic time for the onset of the resonance is proportional
to the square of the pumping field Rabi frequency divided by the excited
state relaxation rate. Further investigation of EIT in\ $\Lambda $ systems
involving the two hyperfine levels of alkaline atoms, can be found in \cite
{KORSUNSKY97} for different pump intensity regimes. In this work EIT in open 
$\Lambda $ systems is also considered and experimental data presented for $%
Na $. Coherence resonances in open three level systems were also
theoretically and experimentally investigated using the Hanle/EIT scheme 
\cite{RENZONI97,RENZONI98,RENZONI99}. In this scheme, the frequencies of the
optical fields are kept fixed and the atomic energy levels are scanned by a
magnetic field. Analyzing in detail the case of the $F_{g}=1\rightarrow
F_{e}=0$ transition, the authors conclude that the resonance width (in terms
of the tunable magnetic field) decreases without limit as $t^{-1/2}$\cite
{RENZONI99}, $t$ being the interaction time. It is argued that this behavior
is characteristic of open systems. Recently, the transient behavior of EIT
resonances for sudden turn on and off of an intense pump field was
experimentally investigated on a sample of magneto-optically cooled atoms 
\cite{CHEN98,GREENTREE02}. Large attention has also been paid to the
temporal evolution of dark resonances when the light fields are switched on
and off adiabatically\cite{KUKLINSKY89,YELIN00,FLEISCHHAUER00}. In this
case, reversible transfer of information from the light field to the atomic
system can take place in connection with such effects as slow light
propagation \cite{HAU99,KASH99} and light storage \cite{PHILLIPS01}.

In this paper, we explore the evolution of the probe absorption spectrum in
DTLS for atoms having spent a variable amount of time in presence of the
pump and probe fields. Only field intensities below saturation are
considered. Specifically, we focus on the analysis and comparison of the
coherence resonances occurring on the cycling transitions of the $D_{2}$ $%
^{85}Rb$ lines when the atoms are illuminated by a pump and a probe field
with tunable frequency difference and orthogonal linear polarizations. For
the first transition studied, $5S_{1/2}\left( F=2\right) \rightarrow
5P_{3/2}\left( F^{\prime }=1\right) $ of $^{85}Rb$, EIT takes place while
the second $5S_{1/2}\left( F=3\right) \rightarrow 5P_{3/2}\left( F^{\prime
}=4\right) $ transition gives rise to EIA. The transient spectroscopy of the
EIT resonances presented here is complementary of the observations in \cite
{RENZONI97} for the $D_{1}$ line of $Na$ atoms under similar excitation
conditions. The differences between the results in \cite{RENZONI97,RENZONI99}
obtained with a Hanle setup and those presented here for pump-probe
spectroscopy are stressed. To the best of our knowledge, the transient
properties of EIA resonances are addressed in this work for the first time.

The paper is organized as follows. In order to introduce the main features
of the dynamics of coherence resonances, the next section is devoted to the
study of these resonances in simple model systems where EIT and EIA\ occur.
In section III the coherence spectroscopy in realistic atomic systems (with
angular momentum degeneracy) is considered and the transient evolution of
EIT/EIA resonances obtained from numerical calculation. In section IV the
experiments carried on a $^{85}Rb$ atomic beam are presented and discussed.
Section V contains the conclusions.

\section{Model systems.}

\subsection{EIT.}

\begin{figure}[tbp]
\begin{center}
\mbox{\epsfig{file=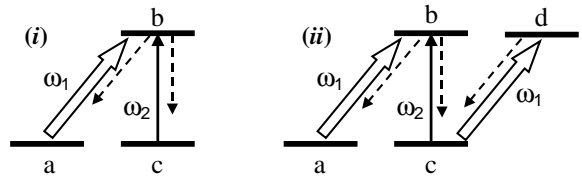,width=3.5in}}
\end{center}
\caption{Model systems for EIT ({\it i}) and EIA ({\it ii}) coherence
resonances. Dashed arrows indicate spontaneous emission decay channels.}
\label{lambdaandn}
\end{figure}

EIT can be conveniently analyzed in the $\Lambda $ system. Consider the
level scheme presented in Fig. \ref{lambdaandn}({\it i}) where the two
degenerate ground levels $a$ and $c$ (same energy) are coupled to the
excited level $b$ (energy $\hbar \omega _{0}$, spontaneous emission decay
rate $\Gamma $) by the pump field ${\cal E}_{1}(t)=2E_{1}\cos \left( \omega
_{1}t\right) $ \ (with $E_{1}$ real) and the probe field ${\cal E}%
_{2}(t)=E_{2}\exp (-i\omega _{2}t)+cc$ respectively. The temporal evolution
of the density matrix of the system is given by \cite{JYOTSNA95}: 
\begin{eqnarray}
\dot{\sigma}_{cc} &=&-i\Omega _{2}\sigma _{bc}+i\Omega _{2}^{\ast }\sigma
_{cb}+\Gamma _{bc}\sigma _{bb}  \nonumber \\
\dot{\sigma}_{bb} &=&-i\Omega _{2}^{\ast }\sigma _{cb}+i\Omega _{2}\sigma
_{bc}+-i\Omega _{1}\left( \sigma _{ab}-\sigma _{ba}\right) -\Gamma \sigma
_{bb}  \nonumber \\
\dot{\sigma}_{aa} &=&-i\Omega _{1}\left( \sigma _{ba}-\sigma _{ab}\right)
+\Gamma _{ba}\sigma _{bb}  \label{blochlambda} \\
\dot{\sigma}_{bc} &=&-\left[ \frac{\Gamma }{2}+i\left( \omega _{0}-\omega
_{2}\right) \right] \sigma _{bc}-i\Omega _{2}^{\ast }\left( \sigma
_{cc}-\sigma _{bb}\right) -i\Omega _{1}\sigma _{ac}  \nonumber \\
\dot{\sigma}_{ab} &=&-\left[ \frac{\Gamma }{2}-i\left( \omega _{0}-\omega
_{1}\right) \right] \sigma _{ab}+i\Omega _{1}\left( \sigma _{aa}-\sigma
_{bb}\right) +i\Omega _{2}\sigma _{ac}  \nonumber \\
\dot{\sigma}_{ac} &=&i\left( \omega _{2}-\omega _{1}\right) \sigma
_{ac}-i\Omega _{1}\sigma _{bc}+i\Omega _{2}^{\ast }\sigma _{ab}  \nonumber
\end{eqnarray}
where the rotating wave approximation was used and the slow variables $%
\sigma _{ii}=\rho _{ii}$ ($i=a,b,c$), $\sigma _{ba}=e^{i\omega _{1}t}\rho
_{ba}$, $\sigma _{bc}=e^{i\omega _{2}t}\rho _{bc}$, $\sigma
_{ac}=e^{i(\omega _{2}-\omega _{1})t}\rho _{ac}$ were introduced. $\Gamma
_{ba}$($\Gamma _{bc}$) is the radiative decay rate from $b$ to $a$($c$) ($%
\Gamma _{ba}+\Gamma _{bc}=\Gamma $), $2\Omega _{1}=\frac{2\mu _{ab}E_{1}}{%
\hbar }$ \ and $2\Omega _{2}=\frac{2\mu _{cb}E_{2}}{\hbar }$ are the Rabi
frequencies for fields ${\cal E}_{1}$ and ${\cal E}_{2}$ respectively ($\mu
_{ab}$ and $\mu _{cb}$ are the electric dipole matrix elements of the
corresponding transitions).

For a weak probe we can seek a solution of Eq. \ref{blochlambda} in the
form: 
\begin{equation}
\sigma (t)\approx \sigma ^{0}(t)+\sigma ^{1}(t)  \label{orders}
\end{equation}
where $\sigma ^{n}(t)$ is of order $n$ in $\Omega _{2}$. For further
simplification let us assume that the pump field is turned on at $t=-\infty $
while the probe field is switched on at $t=0$. With such assumptions, the
zero order term of Eq. \ref{orders} is given, after substitution in Eqs. \ref
{blochlambda}, by $\sigma _{aa}^{0}=\sigma _{bb}^{0}=\sigma _{ba}^{0}=\sigma
_{bc}^{0}=\sigma _{ac}^{0}=0$, $\sigma _{cc}^{0}=1$.

\bigskip
\bigskip
\bigskip
\bigskip

The first order term in Eq. \ref{orders} obeys: 
\begin{eqnarray}
\dot{\sigma}_{bc}^{1} &=&-\left( \frac{\Gamma }{2}-i\delta \right) \sigma
_{bc}^{1}-i\Omega _{2}^{\ast }\sigma _{cc}^{0}-i\Omega _{1}\sigma _{ac}^{1}
\label{firstorderlambda} \\
\dot{\sigma}_{ac}^{1} &=&-i\Omega _{1}\sigma _{bc}^{1}+i\delta \sigma
_{ac}^{1}  \label{caca}
\end{eqnarray}
where $\delta =\omega _{2}-\omega _{1}$ and we assumed for simplicity $%
\omega _{1}=\omega _{0}$ (exact resonance of the pumping field). The initial
condition for the solution of Eqs. \ref{firstorderlambda} is $\sigma
^{1}(0)=0$. Introducing the optical pumping rate $\beta \equiv 2\frac{\Omega
_{1}^{2}}{\Gamma }$, additional simplification of Eqs. \ref{firstorderlambda}
is possible when $\delta ,\beta \ll \Gamma $. In this case, the optical
coherence $\sigma _{bc}^{1}$ adiabatically follows the Raman coherence $%
\sigma _{ac}^{1}$. Taking $\dot{\sigma}_{bc}^{1}=0$ in Eq. \ref
{firstorderlambda} we get: 
\begin{eqnarray}
\sigma _{ac}^{1}\left( t\right) &\simeq &-\frac{2\Omega _{1}\Omega
_{2}^{\ast }\sigma _{cc}^{0}}{\Gamma \left( \beta -i\delta \right) }\left\{
1-\exp \left[ -\left( \beta -i\delta \right) t\right] \right\}
\label{sigmaac} \\
\sigma _{bc}^{1}\left( t\right) &\simeq &-i\frac{\left[ \Omega _{2}^{\ast
}+\Omega _{1}\sigma _{ac}^{1}\left( t\right) \right] }{\left( \frac{\Gamma }{%
2}-i\delta \right) }  \label{sigmabc}
\end{eqnarray}

Consistently with the adiabatic following approximation, the temporal
evolution of optical coherence is dependent on the transient behavior of the
Raman coherence. $\sigma _{ac}^{1}\left( t\right) $ presents a damped
oscillation at frequency $\delta $ with damping coefficient given by the
optical pumping rate $\beta $.

At a given time, the power absorbed from the probe field is: 
\begin{equation}
\alpha _{\Lambda }\left( t\right) \varpropto -%
\mathop{\rm Im}%
\left[ \sigma _{bc}^{1}\left( t\right) \Omega _{2}\right]
\label{probeabstime}
\end{equation}
and its stationary value: 
\begin{equation}
\alpha _{\Lambda }^{ST}\varpropto \left| \Omega _{2}\right| ^{2}%
\mathop{\rm Re}%
\left\{ \frac{1}{\left( \frac{\Gamma }{2}-i\delta \right) }+\frac{2\Omega
_{1}^{2}\sigma _{cc}^{0}}{\Gamma \left( \frac{\Gamma }{2}-i\delta \right)
\left( \beta -i\delta \right) }\right\}  \label{steadyabslambda}
\end{equation}
presents the characteristic EIT dip around $\delta =0$ with a linewidth
(FWHM) of \cite{AGAPEV93}: 
\begin{equation}
\Delta _{\Lambda }\simeq 2\beta \equiv \frac{4\Omega _{1}^{2}}{\Gamma }
\label{deltalambda}
\end{equation}

It is interesting in view of the comparison with the experiments described
below to specify the non-linear absorption defined as the probe absorbed
power in the presence of the pump field minus to the linear absorption (with
no pump field): 
\begin{eqnarray}
\Delta \alpha _{\Lambda }\left( t\right) &\varpropto &%
\mathop{\rm Re}%
\left[ \frac{\sigma _{ac}^{1}\left( t\right) \Omega _{1}\Omega _{2}}{\left( 
\frac{\Gamma }{2}-i\delta \right) }\right]  \label{difprobeabslambda} \\
&=&-KF(\beta /\Gamma ,\delta /\Gamma ,\Gamma t)  \nonumber
\end{eqnarray}
where coefficient $K$ is proportional to the product of the intensities of
the two fields and 
\begin{equation}
F(x,y,\tau )\equiv 
\mathop{\rm Re}%
\left\{ \frac{1-\exp \left[ -\left( x-iy\right) \tau \right] }{\left( \frac{1%
}{2}-iy\right) \left( x-iy\right) }\right\}  \label{functionf}
\end{equation}

\begin{figure}[tbp]
\begin{center}
\mbox{\epsfig{file=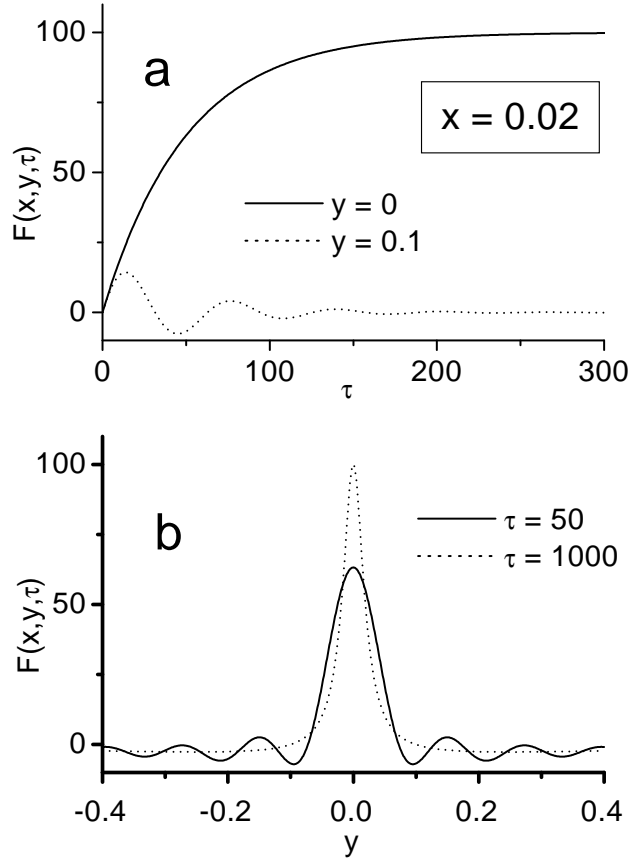,width=3.5in}}
\end{center}
\caption{ Plot of the function $F(x,y,\protect\tau )$ evaluated for $x=0.02$%
. a) $\protect\tau $ dependence for fixed $y$. b) $y$ dependence for fixed $%
\protect\tau $.}
\label{fgraph}
\end{figure}

The main features of the temporal and spectral behavior of the EIT resonance
in the $\Lambda $ system result from the properties of the function $%
F(x,y,\tau )$ which is represented in Fig. \ref{fgraph}. $F(x,y,0)=0$ as
required by the initial condition. For $\tau \gg x^{-1},$ $F(x,y,\infty )$
represent a Lorentzian function of $y$ with FWHM given by $2x$. For $\tau
\ll x^{-1}$, $F(x,y,\tau )$ is an even function of $y$ with a central peak
of width $\sim 4\pi /\tau $ and oscillating wings with period $\sim 2\pi
/\tau $. For fixed $x$ and $y$, $F(x,y,\tau )$ represent a damped
oscillation with frequency\ $y/2\pi $ and damping rate $x$ \cite{KORSUNSKY97}%
.

\subsection{EIA.}

The simplest system presenting EIA is a four-level system in an $N$
configuration [Fig. \ref{lambdaandn}({\it ii})]. EIA appears as the
consequence of coherence transfer from the excited to the lower levels via
spontaneous emission. This system was studied by Taichenachev {\it et al} 
\cite{TAICHENACHEV00} who calculate the analytical expression for the steady
state probe absorption on the $c-b$ transition.

We explore the temporal evolution of the EIA resonance following the same
procedure used for the $\Lambda $ system. The Bloch equations for the system
are initially solved in the presence of the pump field alone and then a
correction of the temporal evolution to first order in the probe field is
determined. Again, for simplicity we assume that the system has reached its
steady state under the presence of the pump field, when the probe field is
turned on at $t=0$. Let the relative electric dipole matrix elements for the
pairs of states $a-b$, $c-b$ and $c-d$ be $A$, $B$ and $1$ respectively with 
$\left| A\right| ^{2}+\left| B\right| ^{2}=1$. Introducing the slowly
varying coefficients: $\sigma _{ii}=\rho _{ii}(i=a,b,c,d)$, $\sigma
_{ba}=e^{i\omega _{1}t}\rho _{ba}$, $\sigma _{bc}=e^{i\omega _{2}t}\rho
_{bc} $, $\sigma _{dc}=e^{i\omega _{1}t}\rho _{dc}$, $\sigma
_{ca}=e^{i(\omega _{1}-\omega _{2})t}\rho _{ca}$, $\sigma _{bd}=e^{i(\omega
_{2}-\omega _{1})t}\rho _{bd}$, $\sigma _{da}=e^{i(2\omega _{1}-\omega
_{2})t}\rho _{da}$, we have in the rotating wave approximation: 
\begin{eqnarray}
\dot{\sigma}_{ac}^{1} &=&i\delta \sigma _{ac}^{1}-i\Omega _{1}A\sigma
_{bc}^{1}+i\Omega _{1}\sigma _{ad}^{1}+\Gamma A\sigma _{bd}^{1}  \nonumber \\
\dot{\sigma}_{bc}^{1} &=&-i\Omega _{1}A\sigma _{ac}^{1}-\left( \frac{\Gamma 
}{2}-i\delta \right) \sigma _{bc}^{1}+i\Omega _{1}\sigma _{bd}^{1}-i\Omega
_{2}^{\ast }\sigma _{cc}^{0}  \label{blochn} \\
\dot{\sigma}_{ad}^{1} &=&i\Omega _{1}\sigma _{ac}^{1}-\left( \frac{\Gamma }{2%
}-i\delta \right) \sigma _{ad}^{1}-i\Omega _{1}A\sigma _{bd}^{1}  \nonumber
\\
\dot{\sigma}_{bd}^{1} &=&i\Omega _{1}\sigma _{bc}^{1}-i\Omega _{1}A\sigma
_{ad}^{1}-\left( \Gamma -i\delta \right) \sigma _{bd}^{1}-i\Omega _{2}^{\ast
}\sigma _{cd}^{0}  \nonumber
\end{eqnarray}
The Rabi frequency $2\Omega _{1}\left( 2\Omega _{2}\right) $ of the
pump(probe) field refers to the $c-d$ ($c-b$) transition and $\sigma
_{cc}^{0}$ and $\sigma _{cd}^{0}$ are the stationary non-zero matrix
elements resulting from the interaction with the pump field alone.

As for the $\Lambda $ system, Eqs. \ref{blochn} can be solved assuming that $%
\sigma _{bc}^1$, $\sigma _{ad}^1$ and $\sigma _{bd}^1$ adiabatically follow
the Raman coherence $\sigma _{ac}^1$. Thus, neglecting $\dot{\sigma}_{bc}^1$%
, $\dot{\sigma}_{ad}^1$ and $\dot{\sigma}_{bd}^1$ in Eqs. \ref{blochn},
after some calculation we get to the lowest order in $\delta $ and $\Omega
_1^2/\Gamma $:

\begin{eqnarray}
\sigma _{ac}^{1}\left( t\right) &\simeq &-\frac{2A\Omega _{1}\Omega
_{2}^{\ast }}{\Gamma \left( \beta ^{\prime }-i\delta \right) }\left\{ 1-\exp %
\left[ -\left( \beta ^{\prime }-i\delta \right) t\right] \right\}
\label{eitramancoherence} \\
\beta ^{\prime } &=&\frac{2\Omega _{1}^{2}}{\Gamma }\left( 1-\left| A\right|
^{2}\right) =\beta \left( 1-\left| A\right| ^{2}\right)
\label{narrowing factor}
\end{eqnarray}
and 
\begin{gather}
\sigma _{bc}^{1}\left( t\right) \simeq \frac{-i\Omega _{2}^{\ast }}{\frac{%
\Gamma }{2}-i\delta }(1-4\frac{\left| \Omega _{1}\right| ^{2}}{\Gamma ^{2}}
\label{sigma1bc} \\
+2A^{2}\frac{\left| \Omega _{1}\right| ^{2}}{\Gamma \left( \beta ^{\prime
}-i\delta \right) }\left\{ 1-\exp \left[ -\left( \beta ^{\prime }-i\delta
\right) t\right] \right\} )  \nonumber
\end{gather}

Using Eq. \ref{probeabstime}, the steady state probe absorbed power is \cite
{TAICHENACHEV00}: 
\begin{equation}
\alpha _{N}^{ST}\varpropto \left| \Omega _{2}\right| ^{2}%
\mathop{\rm Re}%
\left\{ \frac{1}{\frac{\Gamma }{2}-i\delta }\left[ 1-4\frac{\left| \Omega
_{1}\right| ^{2}}{\Gamma ^{2}}+2A^{2}\frac{\left| \Omega _{1}\right| ^{2}}{%
\Gamma \left( \beta ^{\prime }-i\delta \right) }\right] \right\}
\label{steadyn}
\end{equation}
and the time dependent non-linear absorption: 
\begin{equation}
\Delta \alpha _{N}\left( t\right) =-\frac{K^{\prime }}{\left[ \left( \frac{%
\Gamma }{2}\right) ^{2}+\delta ^{2}\right] }+K^{\prime }\frac{A^{2}}{\Gamma
^{2}}F(\beta ^{\prime }/\Gamma ,\delta /\Gamma ,\Gamma t)  \label{nonlinearn}
\end{equation}
with $K^{\prime }$ proportional to the product of the two fields
intensities. The first term in Eq. \ref{nonlinearn} correspond to a broad
(linewidth $\sim \Gamma $) reduction in the absorption due to the saturation
of the $c-d$ transition by the pump field. The second term represents the
EIA resonance. It has a similar spectral and temporal dependence than the
EIT resonance (see Eq. \ref{difprobeabslambda}) except for the sign
inversion. Notice however the reduction by the ``narrowing factor'' $%
1-\left| A\right| ^{2}$ (see Eq. \ref{narrowing factor}) of the damping rate
(and stationary spectral width) of the EIA resonance compared to the EIT
resonance in the $\Lambda $ system. As a consequence, EIA is expected to be
considerably slower in its evolution than EIT for optical transition with
comparable electric dipole matrix elements (comparable Clebsh-Gordan
coefficients for DTLS).

\section{Realistic atomic systems.}

The previous analysis apply to ideal $\Lambda $ and $N$ \ level
configurations. Actual experiments are frequently carried on DTLS where,
depending on their polarization, the pump and probe fields couple different
pairs of Zeeman sublevels resulting in a large variety of (generally more
complex) level schemes.

Before considering in more detail DTLS, let us briefly mention the case of a
pure two level transition driven by a pump and a probe fields of frequencies 
$\omega _{1}$ and $\omega _{2}$ respectively. Unlike the $\Lambda $ and $N$
systems where each field drives different transitions, here the same
transition is driven by both fields. This results in a pulsation of the
populations of the lower and upper levels at the beat frequency $\delta
\equiv \omega _{2}-\omega _{1}$. As a consequence, the atomic coherence
between the ground and excited state posses several frequency components ($%
\omega _{1},\omega _{2},2\omega _{1}-\omega _{2}$) giving rise to a
modulation of the probe absorption at the harmonics of the beat frequency 
\cite{LETOKHOV77}.

\begin{figure}[tbp]
\begin{center}
\mbox{\epsfig{file=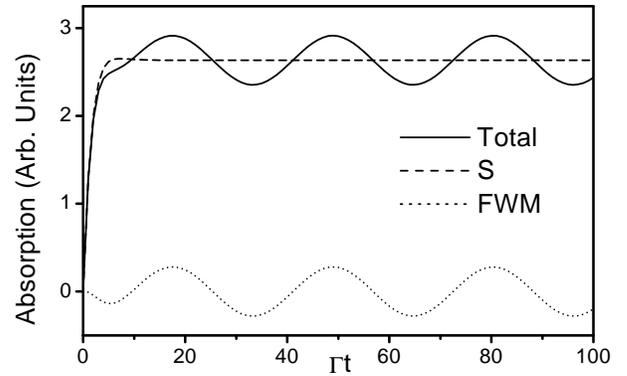,width=3.5in}}
\end{center}
\caption{Transient evolution of the probe absorption of a pure two-level
system. $\Omega _{1}=0.08\ \Gamma $, $\protect\delta =0.1\ \Gamma $. S:
synchronous contribution, FWM: four-wave mixing contribution (see text).}
\label{tlawfwm}
\end{figure}

A similar behavior occur for DTLS driven by two fields. Except for the
simplest combinations of pump and probe polarizations (like $\sigma ^{+},\
\sigma ^{-}$), the excitation of a DTLS with two optical fields result in
population pulsation and the presence of a new frequency component $(2\omega
_{1}-\omega _{2})$ for the atomic coherence between ground and excited state
(optical coherence). We shall call this component ``FWM component'' since it
is responsible for the emission of a new field at frequency $2\omega
_{1}-\omega _{2}$ via four-wave mixing \cite{LEZAMA00}. The FWM component of
the atomic optical coherence results in a modulated contribution to the
probe absorption with frequency $2\delta $. In contrast, we shall call
``synchronous'' (S) the contribution to the probe absorption of the atomic
optical coherence oscillating at frequency $\omega _{2}$. Fig. \ref{tlawfwm}
illustrate the temporal behavior of the probe absorption (for a pure two
level system) showing the S and FWM contributions.

Two features characterize the oscillation due to the FWM contribution: $i)$
It is a permanent oscillation not to be mistaken with the damped oscillation
at frequency $\delta $ that can be present in the S contribution (see Eqs. 
\ref{difprobeabslambda} and \ref{nonlinearn}). $ii)$ The phase of the
oscillation depend on the relative phase between the pump and probe fields 
\cite{KORSUNSKY99}. Both features are of crucial importance in the
experiments. In the experiment described below, only the S contribution to
the probe absorption is observed. The oscillating character of FWM results
in the averaging to zero of this contribution when $\delta \gg T^{-1}$ ($T$
is the temporal resolution in the experiment). Even for $\delta \lesssim
T^{-1}$ the FWM contribution is washed out due to the occurrence of random
(and rapid) fluctuations of the phase difference between the pump and probe
field (due to mechanical vibrations in the setup). The observation of the
FWM contribution should require a careful interferometric control of the
phase difference between the two fields and temporal resolution better than $%
\delta ^{-1}$.

We will next present a theoretical model allowing the calculation, to first
order in the probe field, of the transient probe absorption for DTLS with
arbitrary choices of the ground and excited levels angular momenta and
polarizations of the pump and probe fields. The model is a direct
application to the time domain of the treatment previously used for the
study of the steady state spectral characteristics of coherence resonances 
\cite{LEZAMA00}. It allows the identification of the S and FWM contributions
in the probe absorption.

We consider two degenerate levels: a ground level $g$ of total angular
momentum $F_{g}$ and energy $\hbar \omega _{g}$ and an excited state $e$ of
angular momentum $F_{e}$ and energy $\hbar \omega _{e}$. The radiative
relaxation coefficient of level $e$ is $\Gamma $. We restrict ourselves to
closed transitions. Extension of the model to open transitions is
straightforward \cite{LEZAMA00}.

The atoms are initially in the ground state described by an isotropic
density matrix $\rho _{0}$. They are submitted to a magnetic field $B$ along
the quantization axis and to two classical optical fields:

\begin{quotation}
\[
{\cal E}_{j}(t)=E_{j}\hat{e}_{j}e^{i\omega _{j}t}+E_{j}^{\ast }\hat{e}%
_{j}^{\ast }e^{-i\omega _{j}t},\ (j=1,2) 
\]
$\widehat{e}_{j}$ are complex polarization vectors. The Hamiltonian of the
system is: 
\begin{equation}
H(t)=H_{0}+V_{1}\left( t\right) +V_{2}\left( t\right)  \label{hamtot}
\end{equation}
with:
\end{quotation}

\begin{eqnarray}
H_{0} &=&H_{A}+H_{B}  \label{hb} \\
H_{A} &=&\hbar \left( P_{e}\omega _{e}+P_{g}\omega _{g}\right) \\
H_{B} &=&-\mu _{B}F_{z}\left( g_{g}P_{g}+g_{e}P_{e}\right) B \\
V_{j} &=&E_{j}\hat{e}_{j}\cdot \vec{D}_{ge}e^{i\omega _{j}t}+E_{j}^{\ast }%
\hat{e}_{j}^{\ast }\cdot \vec{D}_{eg}e^{-i\omega _{j}t}  \label{atomfield}
\end{eqnarray}
here $P_{g}$ and $P_{e}$ are the projectors on the ground and excited
manifolds respectively. $H_{B}$ is the Zeeman Hamiltonian, $g_{g}$ and $%
g_{e} $ are the gyromagnetic factors of the ground and excited levels
respectively, $\mu _{B}$ is the Bohr magnetron and $\hbar F_{z}$ the
projection of the total angular momentum along the quantization axis. $\vec{D%
}_{ge}=\vec{D}_{eg}^{\dagger }=P_{g}\vec{D}P_{e}$ is the lowering part of
the atomic dipole operator (we assume that $P_{g}\vec{D}P_{g}=P_{e}\vec{D}%
P_{e}=0$). In Eq. \ref{atomfield} the usual rotating wave approximation is
used.

The temporal evolution of the atomic density matrix $\rho $ is governed by
the master equation:

\begin{eqnarray}
\frac{\partial \rho }{\partial t} &=&-\frac{i}{\hbar }\left[ H,\rho \right] -%
\frac{\Gamma }{2}\left\{ P_{e,}\rho \right\}  \nonumber \\
&&+\Gamma \sum_{q=-1,0,1}Q_{ge}^{q}\rho Q_{eg}^{q}  \label{mastereq}
\end{eqnarray}
where $Q_{ge}^{q}=Q_{eg}^{q\dagger }\ \left( q=-1,0,1\right) $ are the
standard components of the dimensionless operator:

$\smallskip $%
\begin{equation}
\vec{Q}_{ge}=\sqrt{2F_{e}+1}\frac{\vec{D}_{ge}}{\left\langle g\Vert \vec{D}%
\Vert e\right\rangle }  \label{defq}
\end{equation}
$\left\langle g\Vert \vec{D}\Vert e\right\rangle $ is the reduced matrix
element of the electric dipole operator.

In order to find the response of the system to the two fields, we first
consider the effect of the pump field ${\cal E}_{1}$ and next we calculate
the effect of the probe field ${\cal E}_{2}$ to first order.

It is convenient to introduce the slowly varying matrix $\sigma _{0}$ given
by: 
\begin{eqnarray}
\sigma ^{0} &=&\sigma _{gg}^{0}+\sigma _{ee}^{0}+\sigma _{ge}^{0}+\sigma
_{eg}^{0}  \nonumber \\
\sigma _{gg}^{0} &=&P_{g}\rho P_{g}  \nonumber \\
\sigma _{ee}^{0} &=&P_{e}\rho P_{e}  \label{sigma0} \\
\sigma _{ge}^{0} &=&P_{g}\rho P_{e}e^{-i\omega _{1}t}  \nonumber \\
\sigma _{eg}^{0} &=&P_{e}\rho P_{g}e^{i\omega _{1}t}  \nonumber
\end{eqnarray}
after substitution into Eq. \ref{mastereq} (with $V_{2}=0$) one has for $%
\sigma ^{0}(t)$ the equation: 
\begin{eqnarray}
\frac{d\sigma ^{0}}{dt} &=&-\frac{i}{\hbar }\left[ H_{0}+\bar{V}_{1}-\hbar
\omega _{1}P_{e},\sigma ^{0}\right] -\frac{\Gamma }{2}\left\{ P_{e},\sigma
^{0}\right\}  \nonumber \\
&&+\Gamma \sum_{q=-1,0,1}Q_{ge}^{q}\sigma ^{0}Q_{eg}^{q}  \label{pumpeq}
\end{eqnarray}
with: 
\begin{equation}
\bar{V}_{1}=E_{1}\hat{e}_{1}\cdot \vec{D}_{ge}+E_{1}^{\ast }\hat{e}%
_{1}^{\ast }\cdot \vec{D}_{eg}  \label{fieldslow}
\end{equation}
and the initial condition $\sigma ^{0}(0)=\rho _{0}$.

To include the effect (to first order) of the probe field on the atom+pump
system, we seek a solution of Eq. \ref{mastereq} under the form: 
\begin{eqnarray}
\rho _{gg}(t) &=&P_{g}\rho \left( t\right) P_{g}=\sigma _{gg}^{0}+\sigma
_{gg}^{+}e^{i\delta t}+\sigma _{gg}^{-}e^{-i\delta t}  \nonumber \\
\rho _{ee}(t) &=&P_{e}\rho \left( t\right) P_{e}=\sigma _{ee}^{0}+\sigma
_{ee}^{+}e^{i\delta t}+\sigma _{ee}^{-}e^{-i\delta t}  \nonumber \\
\rho _{ge}(t) &=&P_{g}\rho \left( t\right) P_{e}=e^{i\omega _{1}t}\left(
\sigma _{ge}^{0}+\sigma _{ge}^{+}e^{i\delta t}+\sigma _{ge}^{-}e^{-i\delta
t}\right)  \label{firstorder} \\
\rho _{eg}(t) &=&P_{e}\rho \left( t\right) P_{g}=e^{-i\omega _{1}t}\left(
\sigma _{eg}^{0}+\sigma _{eg}^{+}e^{i\delta t}+\sigma _{eg}^{-}e^{-i\delta
t}\right)  \nonumber
\end{eqnarray}
with $\delta =\omega _{2}-\omega _{1}$.

Introducing the non Hermitian matrix: 
\begin{equation}
\sigma =\left( 
\begin{array}{ll}
\sigma _{gg}^{+} & \sigma _{ge}^{+} \\ 
\sigma _{eg}^{+} & \sigma _{ee}^{+}
\end{array}
\right)  \label{nonhermitiansigma}
\end{equation}

After substitution into Eq. \ref{mastereq} and keeping only terms to first
order in $E_{2}$ one has: 
\begin{gather}
\frac{d\sigma }{dt}=-\frac{i}{\hbar }\left[ H_{0}+\bar{V}_{1}-\hbar \omega
_{1}P_{e},\sigma \right] -i\delta \sigma -\frac{\Gamma }{2}\left\{
P_{e},\sigma \right\}  \nonumber \\
+\Gamma \sum_{q=-1,0,1}Q_{ge}^{q}\sigma Q_{eg}^{q}-i\left[ \frac{\Theta _{2}%
}{2},\sigma ^{0}(t)\right]  \label{eqfirstorderdtls}
\end{gather}

where we have introduced $\Theta _{2}\equiv (2\vec{E}_{2}\cdot \vec{D}%
_{ge})/\hbar $. Here the initial condition is: $\sigma (0)=0$.

If the pump field is turned on long before the probe field, then $\sigma
^{0}(t)$ is time independent and Eq. \ref{eqfirstorderdtls} corresponds to a
linear differential equation with constant coefficients. Even if a different
initial condition for the pump field is chosen, the numerical solution of
Eq. \ref{eqfirstorderdtls} can be easily implemented.

The density matrix $\sigma \left( t\right) $ contains all the relevant
information concerning the response to the probe field to first order. The
information concerning the optical atomic polarization oscillating at the
probe frequency (S contribution) is determined by the sub-matrix $\sigma
_{ge}^{+}$ (see Eqs. \ref{firstorder}) while FWM contribution depends on $%
\sigma _{eg}^{+}$.

The probe absorption coefficient due to the S contribution to the atomic
optical polarization is given by: 
\begin{equation}
\alpha _{S}\varpropto \widehat{e}_{2}\cdot Imag\left[ Tr\left( \sigma
_{ge}^{+}\vec{D}_{eg}\right) \right]  \label{abs_S}
\end{equation}
and the absorption dependent on the FWM contribution: 
\begin{equation}
\alpha _{FWM}\varpropto -\widehat{e}_{2}\cdot Imag\left[ Tr\left( \sigma
_{eg}^{+}\vec{D}_{ge}\right) e^{i2\delta t}\right]
\end{equation}

\begin{figure}[tbp]
\begin{center}
\mbox{\epsfig{file=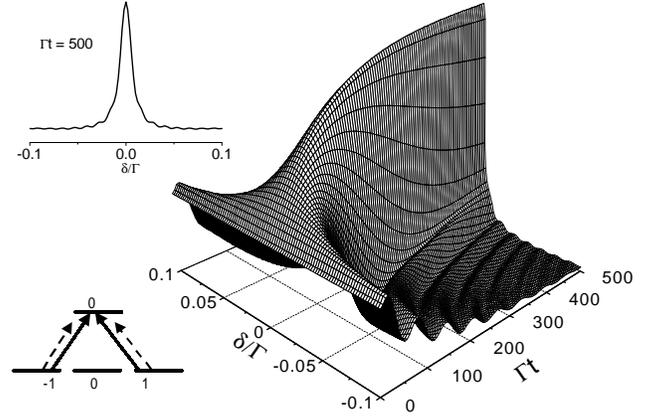,width=3.5in}}
\end{center}
\caption{Calculated temporal and spectral dependence of the nonlinear probe
absorption for a closed $F_{g}=1\rightarrow F_{e}=0$ transition. The pump
and probe polarizations are linear and perpendicular, $\Omega _{1}=0.2\
\Gamma $. Both fields are turned on at $t=0$. The vertical axis (Arb. units)
indicates increasing transparency.}
\label{3deit}
\end{figure}

\begin{figure}[tbp]
\begin{center}
\mbox{\epsfig{file=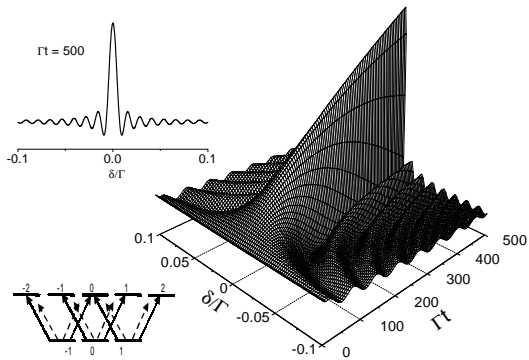,width=3.5in}}
\end{center}
\caption{Calculated temporal and spectral dependence of the nonlinear probe
absorption for a closed $F_{g}=1\rightarrow F_{e}=2$ transition. The pump
and probe polarizations are linear and perpendicular, $\Omega _{1}=0.2\
\Gamma $. Both fields are turned on at $t=0$. The vertical axis (Arb. units)
indicates increasing absorption.}
\label{3deia}
\end{figure}

The predictions of the previous model are illustrated in Figs. \ref{3deit}
and \ref{3deia} showing the time dependence of the absorption spectra
arising from the S contribution for two different transitions with $B=0$.
Only the non-linear probe absorption (total absorption minus linear
absorption) is plotted. The pump and probe fields have linear and
perpendicular polarizations. The calculations were carried in both cases
assuming that the atoms are initially in the ground state (no coherence and
uniform population distribution among Zeeman sublevels) and that the pump
and the probe fields are simultaneously turned on at $t=0$ (as suggested by
the experiments described below). Fig. \ref{3deit} corresponds to an $%
F_{g}=1\rightarrow F_{e}=0$ transition for which EIT occurs. The vertical
axis indicate increasing transparency. The constant level at $t=0$
correspond to the linear absorption for unpumped atoms. Notice the increase
in the average absorption level for nonzero $\delta $ corresponding to the
ground state alignment produced by the pump field. As in the simple models
previously analyzed, the absorption present coherent oscillations at the
Raman detuning $\delta $. At short times, the lineshape is given by a
central peak of width inversely proportional to $t$, with oscillating wings.
At longer times the spectrum approaches a Lorentzian shape with a width
proportional to the intensity of the pumping field \cite{KORSUNSKY97}. The
results presented in Fig. \ref{3deit} correspond to the same atomic system
analyzed in \cite{RENZONI99}. The system is equivalent to an open $\Lambda $
system since the atom can decay from the excited state into the $m_{g}=0$
ground state Zeeman sublevel not coupled by the light. In contrast to what
occurs in the same system for the Hanle type experiment, the width of the
EIT resonance, approaches a finite linewidth at long times determined by the
intensity of the pump field. Such behavior is common to coherence resonances
in closed \cite{LI95} and open \cite{KORSUNSKY97} $\Lambda $ systems
observed through pump-probe spectroscopy. The fact that no qualitative
difference in the time dependence of the resonance width is observed between
closed and open transitions is a consequence of the assumption that the
probe field is weak and does not modify the Zeeman sublevels populations.
Fig. \ref{3deia} corresponds to the EIA\ type transition $F_{g}=1\rightarrow
F_{e}=2$. The vertical axis indicate increasing absorption. The nonlinear
probe absorption was calculated for the same parameters, aside from the
choice of the transition, than in Fig. \ref{3deit}. Observe the noticeable
slowing of the EIA evolution in comparison with EIT. For the time interval
presented in Fig. \ref{3deia} (see inset) the nonlinear absorption spectrum
is still far from reaching the asymptotic Lorentzian shape. This result
suggest that the EIA slowing (and consequently narrowing) introduced above
via the simple $N$ system is a general feature. We have checked numerically
that the characteristic evolution time of the EIA resonances is an
increasing function of the total angular momentum of the atomic levels
involved.

We have chosen to illustrate the preceding analysis,\ the transitions $%
F_{g}=1\rightarrow F_{e}=0$ and $F_{g}=1\rightarrow F_{e}=2$ which are the
simplest (with integer angular momenta) to present EIT and EIA. We have
checked that similar conclusions stand for other choices of the angular
momenta and in particular for the $F_{g}=2\rightarrow F_{e}=1$ and $%
F_{g}=3\rightarrow F_{e}=4$ transitions occurring in the $D_{2}$ line of $%
^{85}Rb$.

Finally, let us briefly comment on the effect of a nonzero magnetic field.
In view of the experimental conditions presented below, we focus on moderate
magnetic fields and restrict the range of variation of $\delta $ in order to
verify $\left| \delta \right| <g_{g}\mu _{B}B<\Gamma $. In consequence, only
the coherence resonance around $\delta =0$ ($\Delta m_{g}=0$) is considered.
We have checked that under such assumptions, the main characteristics
discussed above for the transient spectral behavior of the coherence
resonances are generally preserved \cite{COMMENT}. However, some differences
are observed: An additional frequency component of the nonlinear absorption
oscillating at the Larmor frequency $2g_{g}\mu _{B}B$ appears and there are
also quantitative variations on the resonance peak amplitude and on the
average absorption outside the peak. A detailed study of the influence of
the magnetic field on the coherence resonances dynamics is beyond the scope
of this paper.

\section{Experiment.}

\begin{figure}[tbp]
\begin{center}
\mbox{\epsfig{file=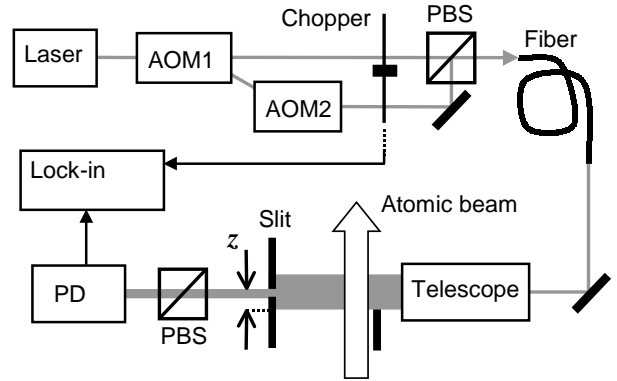,width=3.5in}}
\end{center}
\caption{Experimental setup. PBS: polarizing beam splitter. AOM:
acousto-optic modulator. PD: photodiode.}
\label{setup}
\end{figure}

We have studied the transient spectral behavior of coherence resonances
arising on the $D_{2}$ transitions of $^{85}Rb$ atoms in an atomic beam. The
experimental scheme is represented in Fig. \ref{setup}. The atomic beam has
a $Rb$ reservoir chamber heated to $180{{}^{o}}C$. The atoms exit the
heating chamber through a $1\ cm$ diameter, $6\ mm$ long cylindrical nozzle
formed by an array of parallel tubes of $0.5\ mm$ internal diameter. The
collimation of the atomic beam is achieved with a rectangular slit ($1\
mm\times 10\ mm$) situated $25\ cm$ downstream. The estimated atomic flux
after the collimation slit was $2\times 10^{15}$ $atoms/s$.

The optical beams are obtained by an injected diode laser ($\sim 1\ MHz$
linewidth). Using two consecutive acousto-optic modulators (AOM), one of
which is driven by a tunable RF source, mutually coherent pump and probe
beams are obtained \cite{AKULSHIN98} with variable frequency offset. The two
optical fields have orthogonal linear polarizations. They are combined in a
polarizing beam splitter and sent though a $50\ cm$ long single-mode fiber
(that preserves polarization) for perfect overlap. After the fiber, the beam
is expanded with a telescope. Only the central part of the expanded beam
where the intensity is approximately constant (less than $10\%$ variation)
is used to illuminate the atomic sample.

The atomic beam is intersected at right angle by the light beam. The shadow
of the rectilinear edge of an opaque screen introduces a sudden transition
between obscurity and light for the incoming atoms. Using a movable slit of
width $\Delta z=0.5\ mm$\ wide, the transmitted light is collected at
different positions after the entrance of the atoms in the illuminated
region. The probe absorption spectra are recorded as a function of the
frequency difference $\delta $ between the probe and pump fields for
different values of the slit position $z$. Each spectrum corresponds to
atoms having interacted with the light during an average interval $t=z/\bar{v%
}$ where $\bar{v}$ is the mean velocity in the atomic beam. After the slit,
the pump field is blocked with a linear polarizer. The transmitted probe
light is detected with a photodiode.

A small magnetic field ($\sim 2\ G$), oriented in the direction of the light
propagation, was applied to the interaction region to eliminate the
broadening of the coherence resonances arising from magnetic field
inhomogeneities. Due to this field, only $\Delta m_{g}=0$ resonances are
observed which are insensitive to the magnetic field strength \cite
{AKULSHIN98}. The magnetic field introduces a rapid oscillation of the probe
absorption (at the Larmor frequency) with a period of approximately $0.5\
\mu s.$ Since the temporal resolution in the experiment is $\Delta t\simeq
\Delta z/\bar{v}\approx 1.3\ \mu s$ such oscillation is not observed.

In order to improve sensitivity and have direct access to the nonlinear
absorption, a two-frequencies modulation technique was used. The pump and
probe fields are mechanically chopped at frequencies $f_{1}$ and $f_{2}$
respectively and the photodiode current is analyzed with a lock-in amplifier
at the sum frequency $f_{1}+f_{2}$ ($\sim 300\ Hz$).

\begin{figure}[tbp]
\begin{center}
\mbox{\epsfig{file=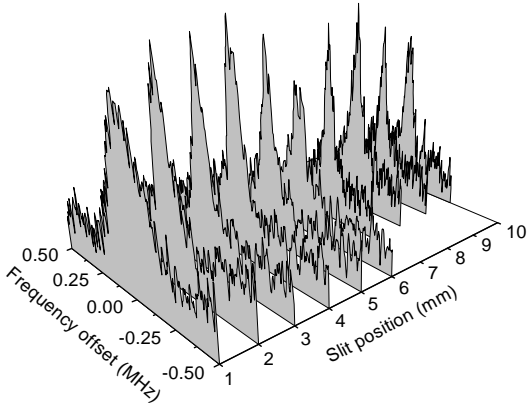,width=3.5in}}
\end{center}
\caption{Experimental nonlinear probe absorption spectra at different
positions of the movable slit for the $5S_{1/2}\left( F=2\right) \rightarrow
5P_{3/2}\left( F^{\prime }=1\right) \ $transition of $\ ^{85}Rb$. The
vertical axis (Arb. units) indicates {\em decreasing} absorption. }
\label{expeit}
\end{figure}

\begin{figure}[tbp]
\begin{center}
\mbox{\epsfig{file=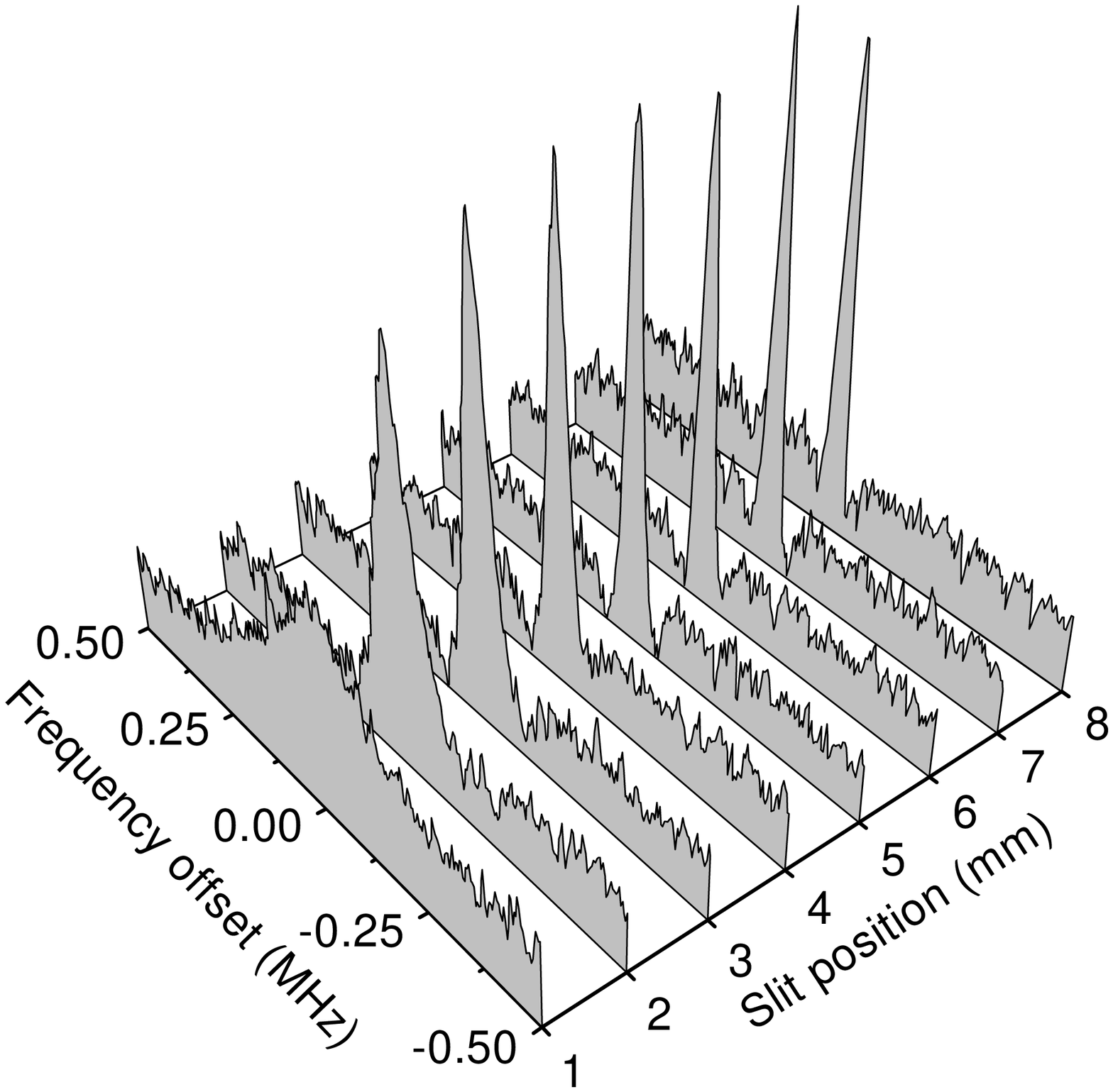,width=3.5in}}
\end{center}
\caption{Experimental nonlinear probe absorption spectra at different
positions of the movable slit for the $5S_{1/2}\left( F=3\right) \rightarrow
5P_{3/2}\left( F^{\prime }=4\right) \ $transition of $\ ^{85}Rb$. The
vertical axis (Arb. units) indicate increasing absorption.}
\label{expeia}
\end{figure}

The temporal and spectral evolution of the probe absorption for the two
closed transitions of the $D_{2}$ line of $^{85}Rb$ are shown in Figs. \ref
{expeit} and \ref{expeia} recorded for a pump field intensity $I_{1}=0.1\
mW/cm^{2}$. Fig. \ref{expeit} correspond to the nonlinear probe absorption
on the $5S_{1/2}\left( F=2\right) \rightarrow 5P_{3/2}\left( F^{\prime
}=1\right) $ transition for which EIT occurs. The vertical axis is linear
and corresponds to increasing transparency. A noticeable the narrowing of
the spectra occurs for the two initial values of the slit position $z$ and
no further narrowing of the spectra is observed for larger values of $z$
(the lineshape is approximately Lorentzian for all $z$'s). This fact,
together with the rapid increase of the peak transparency indicate that the
atoms attain their steady state after $\sim 1\ mm$ flight inside the light
beam. For longer interaction times the linewidth remains unchanged and is
determined by the intensity of the pump field.

\begin{figure}[tbp]
\begin{center}
\mbox{\epsfig{file=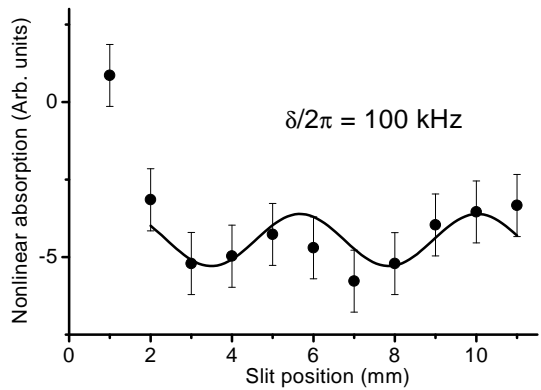,width=3.5in}}
\end{center}
\caption{Time dependence of the nonlinear probe absorption for the $%
5S_{1/2}\left( F=3\right) \rightarrow 5P_{3/2}\left( F^{\prime }=4\right) \ $%
transition of $\ ^{85}Rb$ for $\protect\delta /2\protect\pi =100\ kHz$.
Solid line: sine wave fit.}
\label{oscillation}
\end{figure}

Fig. \ref{expeia} \ corresponds to the nonlinear probe absorption on the $%
5S_{1/2}\left( F=3\right) \rightarrow 5P_{3/2}\left( F^{\prime }=4\right) $
transition (EIA\ resonance) recorded for the same pump intensity. The
vertical axis corresponds to increasing absorption. A much slower evolution
is observed in comparison with the EIT resonance. Notice the continuous
narrowing of the coherence resonance for increasing values of $z$ over all
the range shown and the relatively slow increase of the absorption peak for $%
\delta =0$ as a function of $z$. Even for the largest $z$, the lineshape is
clearly not Lorentzian. It resembles the characteristic shape of the
function $F(x,y,\tau )$ for $\tau \lesssim x^{-1}$ \ and is in good
correspondence with the calculated lineshape (see the inset in Fig. \ref
{3deia}). The narrowest peak observed for EIA \ has approximately $20\ kHz$
FWHM. Fig. \ref{oscillation} illustrate the oscillatory behavior of the
absorption as a function of $z$ for $\delta /2\pi =100\ kHz$. The spatial
frequency of a sine wave fitted to the experimental data (solid line in Fig. 
\ref{oscillation}) is consistent with a mean atomic velocity $\bar{v}\simeq
4.3\pm 0.4\times 10^{2}\ m/s$ in agreement with the value estimated from the 
$Rb$ reservoir temperature. One concludes that in the conditions of Fig. \ref
{3deia}, the steady state is not reached within the observation range.
Considering that the line strength of the $5S_{1/2}\left( F=3\right)
\rightarrow 5P_{3/2}\left( F^{\prime }=4\right) $ transition is larger by a
factor of $3$ than that of the $5S_{1/2}\left( F=2\right) \rightarrow
5P_{3/2}\left( F^{\prime }=1\right) $ transition, one has to conclude that
the slower EIA evolution (compared to EIT) is not due to a smaller pump
field Rabi frequency. Instead the observation suggest that it is due to a
``narrowing factor'' as the one introduced in Eq \ref{narrowing factor}.

\section{Conclusions.}

We have examined the temporal evolution of the spectral response of
degenerate two level system observed using pump probe spectroscopy. The main
features of the transient spectra arise from the consideration of simple
models appropriate to the study of EIA and EIT. These models predict a
similar evolution of the transient response in both cases although EIA
appear to be considerably slower due to the narrowing factor identified in
Eq. \ref{narrowing factor}. A convenient model for the calculation of the
transient response of DTLS was presented. It allows the identification and
separation of the synchronous and the oscillating FWM contributions to the
probe absorption. The numerically calculated transient evolutions of the EIT
and EIA resonances occurring for $F_{g}=1\rightarrow F_{e}=0$ and $%
F_{g}=1\rightarrow F_{e}=2$ transitions respectively, are in good
qualitative agreement with the simple models previously discussed. In both
cases, the resonance width decreases as the inverse of the interaction time
and approaches an asymptotic value determined by the pump field intensity.
Under the same excitation conditions the EIA resonances are slower (and
consequently narrower). Good qualitative agreement with the predicted
transient spectral behavior was obtained for the EIT and EIA\ transitions on
the $D_{2}$ line of $^{85}Rb$.

\section{Acknowledgments.}

The authors acknowledge stimulating discussions with D. Bloch. This work was
supported by the Uruguayan agencies:\ CONICYT, CSIC and PEDECIBA and by ECOS
(France).



\begin{references}
\bibitem{COMMENT}  The transition $F_{g}=1\rightarrow F_{e}=0$ is a special
case since for nonzero magnetic field the pump will eventually drive all the
population into the sink state $m_{g}=0$ not coupled by the probe.
\end{references}
\end{document}